\documentclass{article}
\usepackage[accepted]{vietnam} 
\usepackage{natbib}
\usepackage{graphicx}      
\usepackage{wrapfig}
\usepackage{color}
\usepackage{hyperref} %----set up hyper link to citations, equations...
\hypersetup{colorlinks=true,citecolor=blue}
\usepackage{amsmath, amssymb, amsthm}

\def \bea {\begin{eqnarray}}
\def \ena {\end{eqnarray}}                  
\def \bee {\begin{equation}}
\def \ene {\end{equation}}
\def    \simlt  {\lower.5ex\hbox{$\; \buildrel < \over \sim \;$}}
\def    \simgt  {\lower.5ex\hbox{$\; \buildrel > \over \sim \;$}}

\newcommand     \mum    {\,\mu{\rm m}}  % to use in math mode

\usepackage{color}
\usepackage{hyperref} %----set up hyper link to citations, equations...
\hypersetup{colorlinks=true,citecolor=blue}

\def    \apj  		{\rm {ApJ}}
\def    \apjl  		{\rm {ApJL}}
\def    \mnras  		{\rm {MNRAS}}

\def\be{\begin{equation}}
\def\ee{\end{equation}}
\def\bea{\begin{eqnarray}}
\def\eea{\end{eqnarray}}

\begin{document}
\twocolumn[

\title{Quantitative Polarimetry: A Unified Model of Dust Grain Alignment by Magnetic Radiative Torques}
\titlerunning{A Unified Model of Dust Grain Alignment}

\author{Thiem Hoang}{thiemhoang@kasi.re.kr}
\address{Korea Astronomy and Space Science Institute, Daejeon, Korea}
\vskip 0.5cm 
]

\begin{abstract}
Polarization of optical starlight and far-infrared thermal dust emission due to alignment of interstellar grains offers a powerful window to study magnetic fields in the various astrophysical environments, from the diffuse interstellar medium to accretion disks surrounding young stars. Precision cosmology requires accurate model of Galactic dust polarization for the first detection of Cosmic Microwave Background (CMB) B-modes signal. Such an accurate model is only achieved when a quantitative theory of grain alignment that links grain alignment efficiency with local physical conditions and dust properties is developed and tested. In this paper, we review the successful development of such a quantitative alignment theory, focusing on a unified model of grain alignment based on radiative torques and magnetic relaxation for dust grains incorporated with iron inclusions. We discuss the implication of the unified alignment model for interpreting the latest observational results by {\it Planck} and {\it ALMA}.
\end{abstract}
\section{Introduction}

%1. Review talk: 8 pages
%2. Invited talks and focus moderator talks: 6 pages 
%3. Contributed talks & Focus talk : 4 pages and maximum 2 image
%5. Posters : 1 page and maximum 1 image 
%We should ask SOC to review briefly the papers as well.

%About the content:
%  In your proceeding papers, we would like you to emphasize on your perspective and opinion 
% of your study in the field of star formation, as well as the discussion that you feel important 
% during the conference. That would be great if your proceeding can answer some of 
%the following questions:
%  1. Why do you conduct this study ? How is it important to the field ? 
%  2. What are the advantage and disadvantage of the method the author uses? 
%  3. Does the reach the answer that the author wants? if not, what is next?

The discovery of starlight polarization, more than 65 years ago (\citealt{Hall:1949p5890}; \citealt{Hiltner:1949p5851}) demonstrated that interstellar grains are aspherical and aligned with interstellar magnetic fields. The alignment of dust grains opened a new window into studying the magnetic fields, including the magnetic field topology and strength through starlight polarization (\citealt{1951ApJ...114..206D}; \citealt{1953ApJ...118..113C}) and polarized thermal dust emission (\citealt{Hildebrand:1988p2566}), in various astrophysical environments. Moreover, polarized thermal emission from aligned grains is a significant Galactic foreground source contaminating cosmic microwave background (CMB) radiation (\citealt{2009AIPC.1141..222D}; \citealt{2014arXiv1409.5738P}). It is the most critical challenge to CMB B-mode studies as demonstrated by the joint BICEP2/Keck and Planck data analysis (\citealt{Ade:2015ee}).

The problem of grain alignment has proven to be one of the longest standing problems in astrophysics. Over the six decades, a number of grain alignment mechanisms have been proposed (see \citealt{2007JQSRT.106..225L} and \citealt{LAH15} for reviews).  The initial paradigm of grain alignment is based on the paramagnetic relaxation theory (\citealt{1951ApJ...114..206D}). Some substantial extensions or modifications were suggested to the Davis-Greenstein mechanism. However, an alternative alignment paradigm, based on radiative torques (RATs), has now become the favored mechanism to explain and predict grain alignment in various environmental conditions (\citealt{2007MNRAS.378..910L};  \citealt{Hoang:2008gb}; \citealt{{2009ApJ...697.1316H},{2009ApJ...695.1457H}}). We note that this mechanism was first proposed by \citet{1976Ap&SS..43..291D} and subsequently quantified for several irregular grain shapes by \citet{1996ApJ...470..551D,1997ApJ...480..633D} and \citet{2003ApJ...589..289W}. The first analytical model of RAT alignment is presented by \citet{2007MNRAS.378..910L}. 

Observational evidence for RAT alignment is abundant and becoming increasingly available (\citealt{2007ApJ...665..369A}; \citealt{2008ApJ...674..304W}; \citealt{2010ApJ...720.1045A}; \citealt{2011PASJ...63L..43M}; \citealt{2011A&A...534A..19A}; \citealt{2016arXiv160405305R}). { Some fundamental properties of RAT alignment are observationally tested. For instance, the dependence of alignment on anisotropy direction of radiation have been tested and confirmed by observations (\citealt{2010ApJ...720.1045A}; \citealt{2011A&A...534A..19A}; \citealt{2015ApJ...812L...7V}). The loss of grain alignment toward starless cores because of the reduction of radiation intensity is observed by a number of groups (\citealt{2014A&A...569L...1A}; \citealt{Jones:2014fk})}. Evidence for enhancement in grain alignment by pinwheel torques was recently observed (\citealt{2013ApJ...775...84A}) and successfully reproduced by numerical modeling (\citealt{2015MNRAS.448.1178H}). We refer interested readers to two recent reviews for extended discussions of modern understanding of interstellar grain alignment (\citealt{LAH15}; \citealt{Andersson:2015bq}).

Modern cosmology with CMB polarization requires an accurate model of thermal dust polarization to enable reliable separation of galactic foreground contamination from the polarized CMB signal. We believe that an accurate model of polarized dust emission is achieved only when it is based on solid physics of grain alignment and tested physical properties of interstellar dust. Therefore, it is crucial to calculate the degree of grain alignment to be used for modeling of foreground polarization (cf. \citealt{Draine:2009p3780}). In addition, understanding of the grain alignment is necessary for reliably estimating magnetic fields with the Chandrasekhar - Fermi technique (see \citealt{2008ApJ...679..537F} for a newer edition of the technique).

Most of previous works on RAT alignment deal with ordinary paramagnetic grains (\citealt{1997ApJ...480..633D}; \citealt{Hoang:2008gb}; \citealt{{2009ApJ...697.1316H},{2009ApJ...695.1457H}}), which is based on the assumption that iron atoms are distributed diffusely within silicate grains. Numerous observations, including the recent one (\citealt{Altobelli:2016dl}), indicate that interstellar grains likely contain iron inclusions. The presence of iron clusters can significantly enhance the grain magnetic susceptibility, and grains become {\it superparamagnetic} material (\citealt{Jones:1967p2924}). The question of whether iron inclusions occur only in big silicate grains or both big silicate and carbon grains is still uncertain. Moreover, it is also suggested that there might be an additional dust component of grains that contain iron inclusions in addition to paramagnetic silicate and carbonaceous components (\citealt{2016ApJ...825..136D}). {\it Therefore, it is necessary to formulate a quantitative alignment model for grains containing iron clusters.}

The first work that combines the effect of iron inclusions and RATs was presented by \citet{Lazarian:2008fw}. The authors found that the enhanced paramagnetic relaxation can convert a high-J repellor point induced by RATs in to a high-J attractor point. Recently, \citet{2016ApJ...831..159H} conducted an extensive study on {\it Magnetically enhanced RAT alignment}, which we term MRAT mechanism, and demonstrated that grains with iron inclusions can reach perfect alignment. The detailed description of the MRAT mechanism is shown in Section \ref{sec:MRAT}.

{Recent {\it Planck} results (\citealt{2015A&A...576A.104P}) show a high degree of dust polarization, up to $20\%$, for the diffuse interstellar medium (ISM). Inverse modeling of starlight polarization usually requires perfect alignment of silicate grains to reproduce the maximum polarization degree (\citealt{Draine:2009p3780}; \citealt{2014ApJ...790....6H}). This raises some challenge for the traditional RAT mechanism, but this issue can easily be reconciled by efficient alignment of grains with iron inclusions (\citealt{2016ApJ...831..159H}). In particular, far infrared-submm polarization observations (e.g., by ALMA and EVLA) are able to probe very dense regions where dust coagulation can produce very large grains (i.e., up to cm). For such very large grains in high density conditions, ordinary paramagnetic grains are expected to have slow Larmor precession, such that grains tend to align with the anisotropic radiation direction. If grains contain iron inclusions, the magnetic susceptibility is significantly enhanced, leading to stronger coupling to the magnetic field and deeper probe of magnetic fields into the disk. An extended discussion is presented in Section \ref{sec:discuss}.

\section{Modern Theory of Interstellar Grain Alignment}

\subsection{Paramagnetic relaxation alignment for nanoparticles and small grains}
Interstellar dust grains contain iron atoms that are usually assumed to be diffusely distributed throughout the grain, leading to ordinary paramagnetic material. In an external magnetic field, a rotating paramagnetic grain undergoes paramagnetic relaxation, which dissipates the grain rotational energy and results in the alignment of grain angular momentum with the magnetic field (\citealt{1951ApJ...114..206D}). Paramagnetic relaxation initially was proposed to explain the alignment of interstellar large grains. However, latter it was shown that paramagnetic relaxation is not efficient because grains are randomized by thermal collisions. Nevertheless, numerical calculations by \citet{2014ApJ...790....6H} demonstrate that paramagnetic relaxation can only produce low degree of alignment for small grains. Future observations in the UV wavelengths will be valuable to test the paramagnetic alignment of small grains.

For tiny particles or nanoparticles, the Barnett effect acting within rapidly spinning grains, results in {\it resonance paramagnetic relaxation} (\citealt{2000ApJ...536L..15L}). Previous theoretical predictions using the resonance relaxation mechanism show that the polarization degree of spinning dust emission can be up to a few percents around $10$GHz and falls to below $1\%$ at $\nu>20 $GHz (\citealt{2013ApJ...779..152H}). Therefore, this polarized spinning dust emission appears to be a challenge for the detection of CMB B-mode signal, because simulations (\citealt{2016MNRAS.458.2032R}) show that a level of 1$\%$ spinning dust polarization would affect the CMB B-mode detection. The spinning dust polarization is predicted to be within a few percents (\citealt{2013ApJ...779..152H}; \citealt{2016ApJ...824...18H}) if nanoparticles are aligned by resonance paramagnetic relaxation. Recently, it was suggested that quantization effect of vibrational energy within nanoparticles may suppress the energy transfer from the rotational system to the vibrational system, reducing the degree of magnetic alignment (\citealt{2016ApJ...831...59D}). Meanwhile, it was suggested that Faraday braking can facilitate the dissipation of rotational energy of diamagnetic grains that enhances internal alignment (\citealt{2016MNRAS.457.1626P}). 

\subsection{Quantitative Theory: Radiative Torque Alignment for Paramagnetic Grains}\label{sec:RAT}
The key idea of quantitative polarimetry is first to identify what physical parameters the alignment efficiency, $R$, namely Raleigh reduction factor (\citealt{Greenberg:1968p6020}) depends on and calculate the value $R$ vs. grain size for arbitrary local physical conditions. Next, modeling of dust polarization using $R$ is carried out for realistic environments, and the results can be directly tested with observations. After a long history of more than six decades, a quantitative theory of grain alignment is eventually achieved (see \citealt{LAH15} for a complete review). Below, we review first the classical alignment mechanism and then the modern quantitative alignment model based on radiative torques.

The quantitative study of RAT alignment for ordinary paramagnetic grains was initiated in a series of papers, starting with Lazarian and Hoang (\citealt{2007MNRAS.378..910L}, hereafter LH07) where an analytical model (AMO) of RAT alignment was introduced. In the AMO, the radiative torques are approximately described by analytical torque components $Q_{e1}$ and $Q_{e2}$, which are the torque components parallel and perpendicular to the anisotropic radiation direction. The alignment of paramagnetic grains is dominated by radiative torques, and can be described by the ratio $q^{\max}={\max}(Q_{e1})/{\max}(Q_{e2})$. 

The analytical model was the basis for further theoretical studies in (\citealt{Lazarian:2007p2442}; \citealt{Lazarian:2008fw}; \citealt{Hoang:2008gb} (HL08); \citealt{{2009ApJ...697.1316H},{2009ApJ...695.1457H}}, hereafter HL09ab). In the case of steady motion where only rotational damping due to gas collisions is considered, (i.e., rotational excitations is disregarded), we found that RATs tend to align ordinary paramagnetic grains at attractor points with low magnitude of angular momentum (so-called low-$J$ attractor points), and/or attractor points with high angular momentum (so-called high-$J$ attractor points). 

The parameter space for radiative alignment was identified in LH07 for the combination of $q^{\max}$ and the angle $\psi$ between the direction of the radiation anisotropy and the magnetic field. There exists a narrow window in which the high-J attractors is present, and the alignment can be perfect. The grains with the parameters outside this parameter space were found to be aligned at the low-J attractor point and were shown to be not perfectly aligned. For this generic feature, the degree of radiative alignment can be modeled by a parameter $f_{\rm hi}$--the fraction of grains aligned at high-J attractor points, which allows quantitative predictions of the polarization by RAT alignment for various astrophysical conditions (\citealt{2014MNRAS.438..680H}; \citealt{2015MNRAS.448.1178H}). This approach allows us to predict the polarization spectrum as a function of local physical parameters (gas density and radiation field), but the link to dust properties (e.g., magnetic properties) is not treated.

The quantitative studies of radiative torques in LH07 and HL08 also show that, in the absence of high-J attractor points, the degree of alignment is unlikely to exceed $30\%$ (HL08). Both pinwheel torques due to H$_2$ formation (HL09a) and iron inclusions (LH07b) are suggested to enhance the radiative alignment efficiency. The radiative alignment also predicts that in the presence of strong anisotropic radiation field, the alignment occurs with respect to the radiation direction if the radiative precession is faster than the Larmor precession around the magnetic field.

\subsection{Quantitative Theory: Magnetic Radiative Torque Alignment for Grains with Iron Inclusions}\label{sec:MRAT}
With the goal to linking the alignment efficiency to the magnetic property of dust grains that contains iron inclusions, \citet{2016ApJ...831..159H} (hereafter HL16) have developed a unified model of grain alignment. This model deals with RAT alignment of grains with enhanced magnetic susceptibility due to iron inclusions, which we term MRAT mechanism. Grains with sufficiently large fraction of iron inclusions can get aligned with high-J attractor points, achieving a high degree of alignment. Using our analytical model of RATs, we derive the critical value of the magnetic relaxation parameter $\delta_{m}=\tau_{\rm gas}/\tau_{\rm mag}$ to produce high-J attractor points as functions of $q^{\max}$ and the anisotropic radiation angle relative to the magnetic field $\psi$, where $\tau_{\rm gas}$ and $\tau_{\rm mag}$ are the gas damping and magnetic relaxation times, respectively.

HL16 find that if about $10\%$ of total iron abundance present in silicate grains are forming iron clusters, it is sufficient to produce high-J attractor points for all reasonable values of $q^{\max}$. Figure \ref{fig:deltam_cri} shows the contours of $\delta_{\rm m,cri}$ required to produce high-J attractor points in the plane of $\psi$ and $q^{\max}$. For a wide range of explored parameters of $q^{\max}$ and $\psi$, high-J attractor points are present for $\delta_{\rm m,cri}\le 10$, except for the case $\psi=45^{\circ}$. For instance, for $\psi=0$, higher $\delta_{m}$ is required to have high-J attractors for smaller $q^{\max}$. It is noted that the condition in Figure \ref{fig:deltam_cri} is applicable to any value of $a$. {Moreover, the result is independent on the anisotropy $\gamma$, provided that $\gamma\ne 0$.}

\begin{figure}
%\centering
%\vspace*{-0.5cm}
\includegraphics[width=0.45\textwidth]{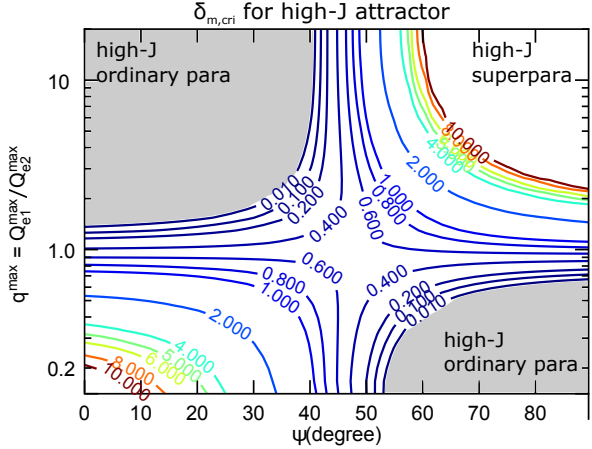}
\caption{Contours of the critical value of magnetic relaxation parameter required to produce high-J attractor points, $\delta_{\rm m, cri}$, in the plane of $\psi$ and $q^{\max}$. Shaded areas indicate the existence of high-J attractor points produced by RATs even for ordinary paramagnetic grains (i.e., no need of magnetic relaxation). White areas indicate the space where the high-J attractor points are only produced in the presence of magnetic relaxation. High-J attractors is universal for $\delta_{\rm m,cri}>10$.}
\label{fig:deltam_cri}
\end{figure}

To calculate the degree of grain alignment as a function of local physical parameters and dust properties, numerical simulations of MRAT alignment are carried out, accounting for stochastic excitations from gas collisions and magnetic fluctuations (HL16). We show that large grains can achieve perfect alignment when the high-J attractor point is present, regardless of the values of $q^{\max}$.  Figure \ref{fig:Ravg_size} shows the dependence of $R$ on the grain size $a$ for different values of $q^{\max}$ at $\psi=0^{\circ}$. It can be seen that the values of $R$ increase with increasing $a$. For the largest magnetic relaxation parameter considered, i.e, $\delta_{\rm m}=30$, the $a\ge 0.1\mu m$ grains are perfectly aligned (upper bound) for all values of $q^{\max}$ considered, whereas smaller grains of $a=0.05$ and $0.08\mu m$ are less efficiently aligned. 

For ordinary paramagnetic case, the values of $R$ are dependent on $q^{\max}$. For $a=0.05-0.08\mu m$, we found that, even $\delta_{\rm m}\gg 1$, those small grains never reach perfect alignment. This can be explained by the fact that the latter, smaller grains are not rotating suprathermally because of weaker RATs. Moreover, for the case of $\psi=60^{\circ}$, we do not have perfect alignment of $a=0.1\mu m$ because those grains only rotate at thermal rates. Furthermore, the parameter space for the dependence of $R$ vs. $a$ is quite broad for the RATs with $q^{\max}=0.5$ and $1$, and it becomes very focused for the two other cases of $q^{\max}=2$ and $4$.

\begin{figure}
%\centering
\includegraphics[width=0.45\textwidth]{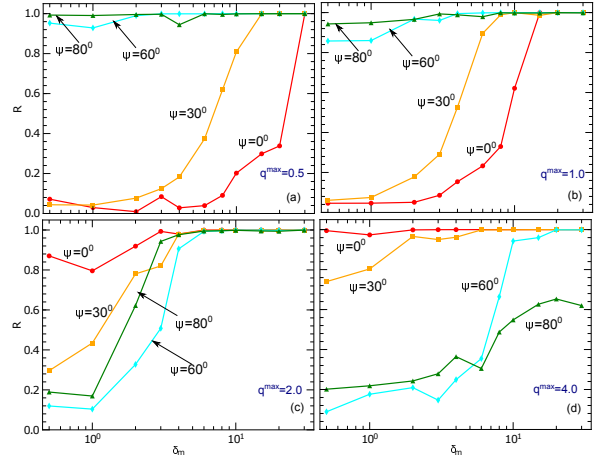}
\caption{Variation of the Rayleigh reduction factor $R$ with $\delta_{\rm m}$ for $\psi=0, 30, 60, 80^{\circ}$ and $q^{\max}=0.5, 1, 2, 4$ (panels (a)-(d)). The MRAT alignment is nearly perfect for the largest $\delta_{\rm m}$ where the alignment has high-J attractor points, such that all grains eventually become aligned. Grain size $a=0.2\mum$ is considered.}
\label{fig:Ravg_delta}
\end{figure}

\begin{figure}
\centering
%\vspace*{0.5cm}
\includegraphics[width=0.43\textwidth]{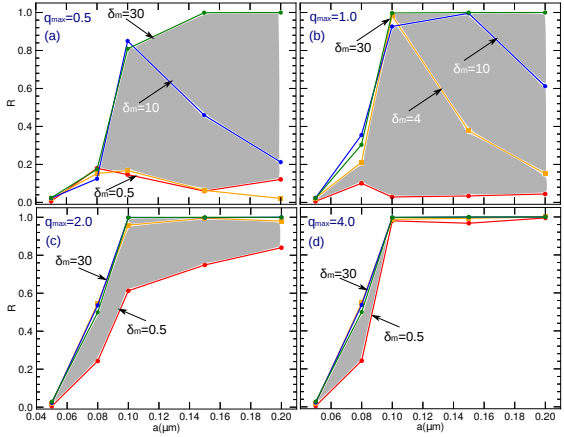}
\caption{Dependence of $R$ on the grain size $a$. Big grains ($a\ge 0.1\mu m$) are mostly perfectly aligned for $\delta_{\rm m}\ge 30$, while grains $a<0.1\mu m$ tend to increase with $a$. The results for $\psi=0^{\circ}$ are shown.}
\label{fig:Ravg_size}
\end{figure}

Figure \ref{fig:Ravg_delta} shows the dependence of $R$ on $\delta_{\rm m}$ for the different values of $q^{\max}$ for grains of size $a=0.2\mum$. For $q^{\max}=0.5$ and $1$, the MRAT is almost perfect for $\psi=60^{\circ}$ and $80^{\circ}$ for all values of $\delta_{\rm m}$, whereas the perfect alignment is only achieved with large value of $\delta_{\rm m}$ for $\psi=0^{\circ}$ and $30^{\circ}$. 
This arises directly from Figure \ref{fig:deltam_cri} that the high-J attractor points are present at $\psi\ge 60^{\circ}$ even with low values of $\delta_{\rm m}$. For $q^{\max}=2$ and $4$, the same happens with $\psi=0^{\circ}$ and $30^{\circ}$. 
For all cases of $q^{\max}$, the MRAT alignment becomes perfect for $\delta_{\rm m}\ge 30$ (panels (a)-(c)). For the case of $q^{\max}=4$, the value ${R}$ for $\psi=80^{\circ}$ cannot reach perfect alignment because grains do not rotate suprathermally at this large angle between the radiation and the magnetic field.

\section{Implications of the Unified Alignment Theory for Observations}\label{sec:discuss}
\subsection{The diffuse interstellar medium}
Recent results by {\it Planck} (\citealt{2015A&A...576A.104P}) show a high polarization degree up to $20\%$ for the diffuse interstellar medium (ISM). Inverse modeling of starlight polarization usually requires perfect alignment of silicate grains to reproduce the maximum possible polarization (\citealt{Draine:2009p3780}; HLM14). This raises some challenge for the classical RAT mechanism described in LH07a. The results there show that the RAT alignment can easily reach perfect if the high-J attractor point is present (also HL08). In contrast, without high-J attractor points, the degree of alignment is only $20-30\%$ (see Figure \ref{fig:Ravg_delta}). This, by itself, does not disqualify RATs from explaining the observed alignment. First of all, the existence of the high-J attractor point depends on the $q^{\max}$ parameter of grains, but its distribution is uncertain for irregular interstellar grains. Moreover, in the absence of high-J attractor points, the degree of alignment still can be significantly increased if grains are also subject to suprathermal torques and the evidence of this effect has been claimed in observations (\citealt{2013ApJ...775...84A}). 

The MRAT scenario (\citealt{2016ApJ...831..159H}}) provides a promising explanation for the perfect alignment  of dust grains, which is supported by the increasing evidence of iron inclusions in big dust grains (e.g., \citealt{Altobelli:2016dl}). If the MRAT scenario is confirmed, this will have important consequences for interpreting the observational data in environment conditions beyond the diffuse ISM, including molecular clouds, dense cores, circumstellar disk, and cometary coma.

\subsection{Probing magnetic fields in protoplanetary disks with aligned grains}

Large grains can be aligned by RATs in accretion disks (\citealt{2007ApJ...669.1085C}). There, however, the observations were not showing good correspondence with theoretical predictions (see \citealt{2009ApJ...704.1204H}). The results were also critically analyzed in a recent study (\citealt{2014MNRAS.438..680H}). Two reasons that are responsible for lower degrees of alignment of grains, including the failure of the internal relaxation to bring the grain angular momentum ${\bf J}$ to align with the axis of the maximum moment of inertia $\hat{\bf a}_{1}$ and the slow Larmor precession of ${\bf J}$ about the magnetic field compared to randomization by gas collisions in the accretion disks. The former effect does not completely suppress the RAT alignment (\citealt{2009ApJ...697.1316H}), but makes the alignment more selective in terms of the grains that can be aligned. The effect of the reduced Larmor precession within very large grains destroys the alignment with respect to the magnetic field. 

Interestingly, recent observations of submm/mm polarization by ALMA reveal strong polarization of thermal dust from disks. It is suggested that the self-scattering by large grains is responsible for such a high polarization (\citealt{2015ApJ...809...78K}), although alignment of large grains is also suggested. Now let us evaluate the maximum size of dust grains that can be aligned with respect to the magnetic field in the light of our unified model of grain alignment. 

Following HL16, the largest grain size that the alignment still occurs with the magnetic field ${\bf B}$ is estimated to be
\bea
a_{\rm ali, B} < 436 \frac{N_{cl,5}\phi_{sp,-2}B_{3}}{n_{10}T_{2}^{1/2}\Gamma_{\|}} \mu m,\label{eq:aLar}
\ena
where $N_{\rm cl}$ is the number of iron atoms per cluster, spanning $\sim 20$ to $10^{5}$ (\citealt{Jones:1967p2924}), $B_{3}=B/1mG$, $n_{10}=n_{\rm H}/10^{10}{\rm cm}^{-3}$, $T_{2}=T_{\rm gas}/100$K, and $\Gamma_{\|}$ is the geometry parameter of unity order. Here, $N_{cl,5}=N_{\rm cl}/10^{5}$, $n_{\rm H}$ and $T_{\rm gas}$ are gas density and temperature.

For accretion disks with $B\sim 10$mG (\citealt{2006Sci...313..812G}) and $T_{\rm gas}\sim 25$K (\citealt{1997ApJ...490..368C}), we expect to have alignment of very big grains with the magnetic field in the disk interior for $a_{ali,B}\sim 218 N_{cl,5}/n_{10}\mu m$. For a typical flared disk around T-Tauri stars, the density at radius $d$ is $n_{\rm H}\sim 10^{9}(d/100AU)^{-39/14}\exp\left(-z^{2}/2H^{2}\right)cm^{-3}$ with $H$ the scale height and $z$ the distance from the disk plane (\citealt{2014MNRAS.438..680H}). Therefore, grains as large as $a_{\rm ali,B}\sim 2$mm can be aligned in the disk-plane at distance $d>100$ AU. At disk radius $d=50$ AU, grains up to $a_{\rm ali,B}\sim 0.5$mm can be aligned with the magnetic field in the disk plane; at height $z\sim 2H$, this upper size increases to $a_{\rm ali,B}\sim 4$mm. We note that a slightly different conclusion is obtained recently (\citealt{Tazaki:2017wb}) because they chose a low value of $N_{\rm cl}=10^{3}$, compared to $N_{\rm cl}=10^{5}$ adopted in HL16.

{For the similar conditions of T-Tauri disks aforementioned, we expect Barnett relaxation is important only for small grains upto $a_{\rm ali,Bar}\sim 0.5\mu m$ at radius $d>100$ AU. Nevertheless, grains with inefficient internal relaxation may still be aligned with the magnetic field by strong RATs (\citealt{2009ApJ...697.1316H}). Therefore, the potential of alignment of the millimeter-sized grains with the magnetic field in the accretion disks remains feasible if grains are incorporated with iron inclusions and the magnetic field strength is about mG. This prediction is recently confirmed by detailed modeling of polarization from aligned grains in accretion disks (\citealt{Tazaki:2017wb}).

\subsection{Towards self-consistent modeling of dust polarization for the CMB B-modes quest}

The development of a unified grain alignment theory has important implications for CMB B-mode studies. Apart from being the dominant source of polarized thermal emission via electric dipole mechanism for the frequencies larger than $300$ GHz, grains with iron inclusions can also be responsible for magnetic dipole emission at lower frequencies (\citealt{2013ApJ...765..159D}; \citealt{2016ApJ...821...91H}). The latter emission mechanism is closely related to magnetic inclusions in grains, which are the inclusions that are invoked for the MRAT alignment. This emission, if its magneto-dipole nature is confirmed, can be used to constrain the abundance and properties of the magnetic inclusions that are appealed to within the MRAT mechanism. 

Polarized thermal dust emission is shown to be the most crucial challenge for detection of CMB B-modes, thus an accurate model of dust polarization is urgently needed. The size-dependence degree of grain alignment is an essential ingredient for modeling the polarization spectrum of thermal dust emission and magnetic dust emission (MDE). We have obtained such a size-dependence degree of alignment from simulations using AMO, which relates directly the degree of alignment with the magnetic susceptibility, radiation fields, and local environmental conditions. It will be used to construct a self-consistent model of magnetic dipole emission and polarization spectrum.

\section{Conclusions and Future Perspectives}
For the first time, a quantitative theory of grain alignment based on radiative torques is developed and tested with observations. Such a theory allows us to predict alignment efficiency as function of dust properties and local physical parameters, which paves the way for self-consistent modeling of multiwavelength dust polarization. This has great significance for precision CMB B-mode studies as well as more realistic understanding of magnetic fields in star formation processes. In particular, iron inclusions in big grains can increase the upper cutoff of grain size that are still coupled to the magnetic field due to fast Larmor precession. This allows using mm-polarimetry (e.g., with ALMA and EVLA) to probe magnetic fields in protoplanetary disks and study its role on planet formation. Nevertheless, to disentangle the polarization from aligned grains from self-scattering in the protoplanetary disks, detailed physical modeling and multi-wavelength observations are needed (\citealt{2015ApJ...809...78K}; \citealt{2016MNRAS.460.4109Y}). 

Although observations establish that iron is depleted into interstellar dust, the exact form of dust-phase iron is still uncertain. Future observational studies should focus on testing whether ion inclusions are incorporated in big dust grains or present as free-flier nanoparticles in the ISM. Interestingly, the latest experiment (\citealt{Anonymous:lUXlSJiR}) reveals that pure iron grains are rare due to low sticky probability, suggesting that iron is more likely present in dust grains as inclusions/compounds instead of being free-fliers. Moreover, dust grains with iron inclusions may be an independent dust component from silicate, carbonaceous grains (\citealt{2016ApJ...825..136D}). We note that the fraction of grains with iron inclusions as well as the level of iron inclusions likely vary in the ISM, thus the perfect alignment is not expected throughout the ISM, resulting in the scatter in the polarization efficiency seen by observations (e.g., {\it Planck} satellite). 

Iron inclusions are also suggested to induce grain alignment in cometary coma as observed by {\it Rosetta} satellite (\citealt{Kolokolova:2016cd}). Similar to RATs, iron inclusions are found to enhance alignment by mechanical torques for some irregular shapes (Hoang, Cho, \& Lazarian, in preparation).  

There are still important questions on grain alignment that need to be addressed. Those include: Are grains with iron inclusions independent of paramagnetic grains? Can carbonaceous grains with iron inclusions be aligned? What is the residual alignment efficiency of nanoparticles/ultrasmall grains? What is the alignment efficiency of very large (mm) grains without internal relaxation?

%\section*{Acknowledgments}

%This is where one places acknowledgments for funding bodies etc. Note that there are no section numbers for the Acknowledgments, Appendix or References.

%\section*{Appendix}

%--------------adding references-----------------------------------
%\bibliographystyle{abntex2-alf}

%\bibliography{/Users/thiemhoang/Dropbox/Papers2/cites_paperApJ,/Users/thiemhoang/Dropbox/Papers2/cites_Books}

%\bibliography{ms1.bbl}

\end{document}